# Modifying Electrochemical Doping in Light-Emitting Electrochemical Cells with Gold Nanoparticles

Ajay K. Poonia[1, 2, 3*], Anton Kirch[2], Joan Ràfols-Ribé[1, 2], Lucrezia Catanzaro[4], Anish Rao[5], Karol Kołątaj[5], Vittorio Scardaci[4], Giuseppe Compagnini[4], Guillermo P. Acuna[5], Ludvig Edman[2, 3*], and Nicolò Maccaferri[1, 3*]

*[1]Ultrafast Nanoscience Group, Department of Physics, Umeå University, Umeå, 901 87, Sweden*

*[2]The Organic Photonics and Electronics Group, Department of Physics, Umeå University, Umeå, 901 87, Sweden*

*[3]Wallenberg Initiative Materials Science for Sustainability, Department of Physics, Umeå University, Umeå, 901 87, Sweden*

*[4]Department of Chemical Sciences, University of Catania, Catania, 95125, Italy*

*[5]Department of Physics, University of Fribourg, Fribourg, 1700, Switzerland*

**ajay.poonia@umu.se*

**ludvig.edman@umu.se*

**nicolo.maccaferri@umu.se*

**Abstract**

Electrochemical doping offers dynamic control of the electronic properties of organic semiconductors, and it is the enabling feature of a range of technologies, including electrochemical transistors, energy-storage devices, light-emitting electrochemical cells (LECs), and bioelectronics. Electrochemical doping is commonly controlled by the selection of the constituents in the active material of the device or the applied voltage bias, but herein we report that the incorporation of Au nanoparticles (Au-NPs) at an electrode interface can constitute an alternative control parameter. The LEC features balanced p- and n-type electrochemical doping that forms a p-n junction doping structure in its active material, and we find that it is possible to reshape this doping profile by incorporating Au-NPs at an electrode interface. Specifically, we establish that the inclusion of neat non-capped Au-NPs at the anodic interface shifts the p-n junction (i.e., the emission zone) away from the anode. In contrast, the inclusion of Au-NPs capped with sodium citrate is found to reverse this behavior, so that the emission zone is instead moved towards the anode. We utilize this control parameter to shift the emission zone towards a position of constructive (destructive) interference, as manifested in a strong increase (decrease) of the LEC emission efficiency. Our findings establish an interfacial strategy for modulating the spatial profile of electrochemical doping and tuning device performance without altering the chemistry of the active material, relying instead on the surface modification of one electrode. This approach is important because it provides a versatile and minimally invasive route to optimize electrochemical devices while preserving the intrinsic properties and formulation of the active material.

## Introduction

The electronic, optical, and mechanical properties of organic semiconductors (OSCs) can be dynamically controlled by electrochemical doping[1–5], which enables a wide range of technologies, such as electrochemical transistors[4,5], light-emitting electrochemical cells (LECs)[6], energy-storage[7], as well as neuromorphic[8,9] and bioelectronic[10] devices. Unlike conventional chemical doping, which fixes the doping species and the doping density during fabrication, electrochemical doping enables in situ modulation of the spatiotemporal doping concentration for desired properties[11]. The ionic species required for electrochemical doping are either supplied from an adjacent electrolyte reservoir or are blended with the OSC in the active material (AM), which is positioned between two (or more) electrodes[11]. In this context, considerable efforts have been devoted to controlling the electrochemical doping process through the design and development of new OSCs and ionic structures[5,6,12–14], or by the introduction of tuned additives[12,15], including dedicated ion-transport molecules[15]. While these strategies have proven effective[8,9,13,16], an additional broadly applicable handle for manipulating electrochemical doping would be desirable. Here, we investigate whether Au nanoparticles (Au-NPs) incorporated at an electrode interface can act as such a control handle. We specifically use the LEC as a model platform and examine how different types and sizes of Au-NPs introduced at the anodic interface influence the electrochemical doping process.

The LEC comprises a blend of an electroluminescent OSC and mobile ions as the AM sandwiched between two electrodes. When a voltage is applied between the two electrodes, electrochemical p-type (n-type) doping can be achieved through the injection of holes (electrons) at the positive anode (negative cathode) and the electrostatically compensatory motion of ions. The mobile ions migrate to the electrode interfaces to form electric double layers (EDLs), which facilitate efficient and balanced electron and hole injection. The remaining mobile ions redistribute to screen the injected electronic charge, leading to the formation of p-type and n-type doping regions. At the p-n junction, that separates the two doping regions, subsequently injected electrons and holes recombine into excitons that can undergo radiative decay; the p-n junction thus defines the emission zone (EZ)[6,12,17–23] (see Fig. 1a). We utilize in-house developed experimental tools[15,24] to show that Au-NPs placed at the anodic interface of an LEC can strongly displace the EZ within the AM, with both the direction and magnitude of the shift being determined by the Au-NP type and size. Specifically, non-capped neat Au-NPs are observed to induce a cathodic displacement of the EZ, with the larger-sized Au-NPs producing the largest shift. By contrast, Au-NPs capped with an insulating sodium citrate layer cause an anodic shift of EZ, presumably because the insulating ligands inhibit hole injection. We finally exploit this interfacial control parameter to reposit the EZ in the AM towards a region of enhanced optical outcoupling, which translates into a marked increase in the LEC emission efficiency.

## Results and Discussion

We have synthesized and investigated three different types and sizes of Au-NPs: 1) neat uncapped Au-NPs with an average diameter of 11 nm (NP11); 2) neat uncapped Au-NPs with an average diameter of 54 nm (NP54); 3) capped Au-NPs coated with an electronically insulating layer of trisodium citrate and featuring an average diameter of 60 nm (CNP60).

Detailed information on the Au-NP synthesis and characterization is provided in Methods and Supplementary Figs. 1-3.

The AM of the selected LEC consists of a blend[25] of the electroluminescent organic semiconductor Super Yellow (SY), the ion-transporting and ion-dissolving compound hydroxyl-capped trimethylolpropane ethoxylate (TMPE-OH), and the salt $KCF_3SO_3$ (see Fig. 1b for the chemical structures). The control-LEC (Fig. 1c) comprises a spin-coated AM layer positioned between a transparent indium tin oxide (ITO) anode on a glass substrate and a reflective aluminum cathode. The three NP-LECs (Fig. 1d) differ from the control-LEC by the presence of Au-NPs, which were spin-coated onto the ITO surface from an Au-NP dispersion before AM deposition. The dry Au-NP surface coverage on the ITO surface was 12% for NP11, 0.2% for NP54 and 0.2% for CNP60, as determined by scanning electron microscopy (see Supplementary Fig. 3). The AM thickness was ~150 nm for all four LEC devices, as shown in Supplementary Fig. 4. Further details about the device fabrication are provided in Methods.

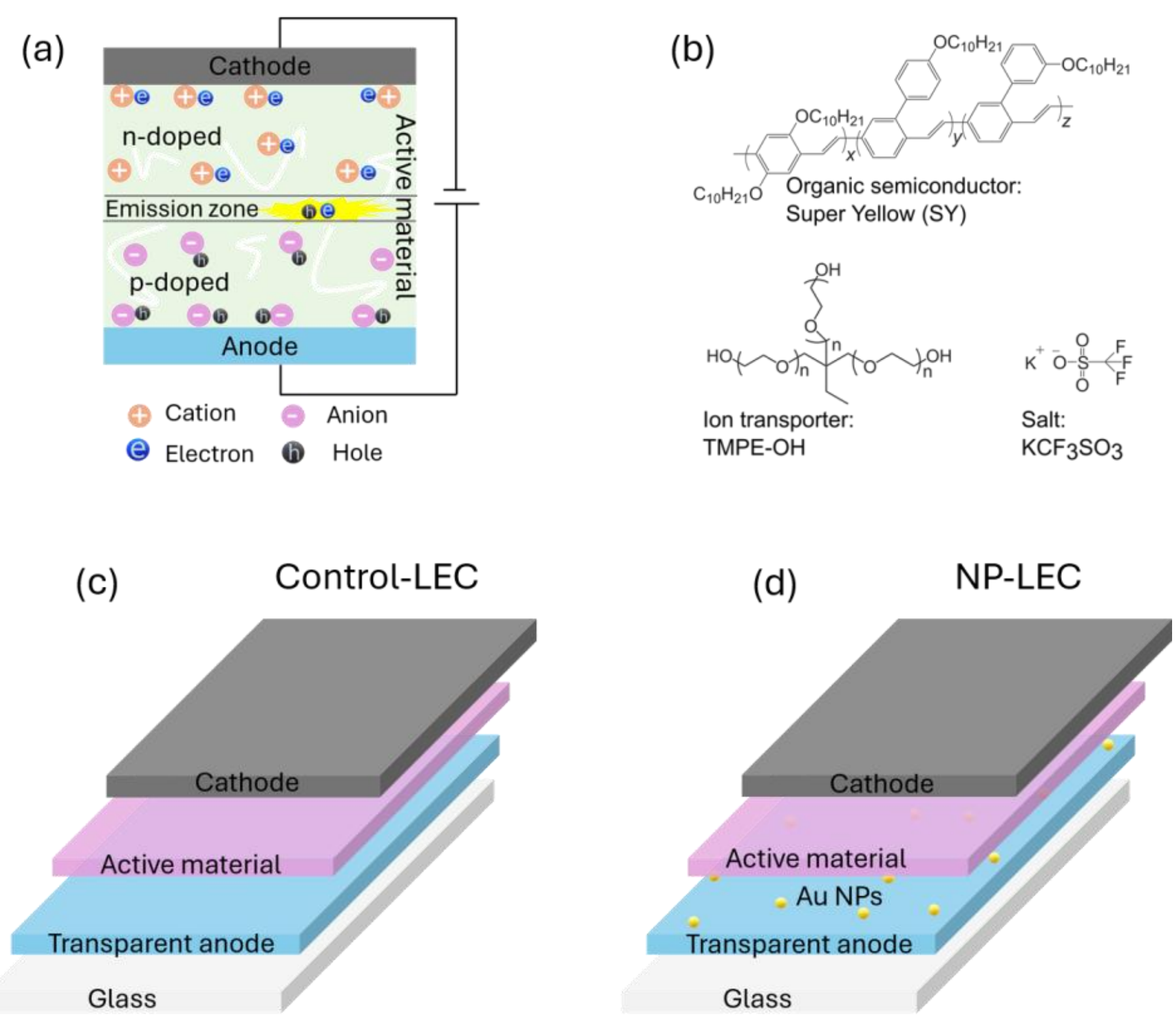

**Figure 1. LEC structure and operation.** (a) Schematic of the p-n junction doping structure that forms in the AM of LEC by electrochemical p- and n-type doping. The light emission originates from the p-n junction that defines the EZ. (b) The chemical structures of the AM components. The device architecture of (c) the control-LEC (d) and the NP-LEC, with the Au-NPs positioned at the anode/AM interface.

In Figs. 2a,b we report the measured transient evolution of the voltage and the luminance, respectively, for the LECs during electric driving by a current density of 25 mA $cm^{-2}$. Additional measurements supporting the reproducibility of the results are shown in Supplementary Fig. 5. All four LEC devices exhibit a decrease in voltage and an increase in luminance during the first few minutes of operation, which is in line with the formation of EDLs and electrochemical doping[26]. Notably, the NP11-LEC and NP54-LEC exhibit significantly enhanced luminance compared to the control-LEC, while the voltage transients are essentially identical.

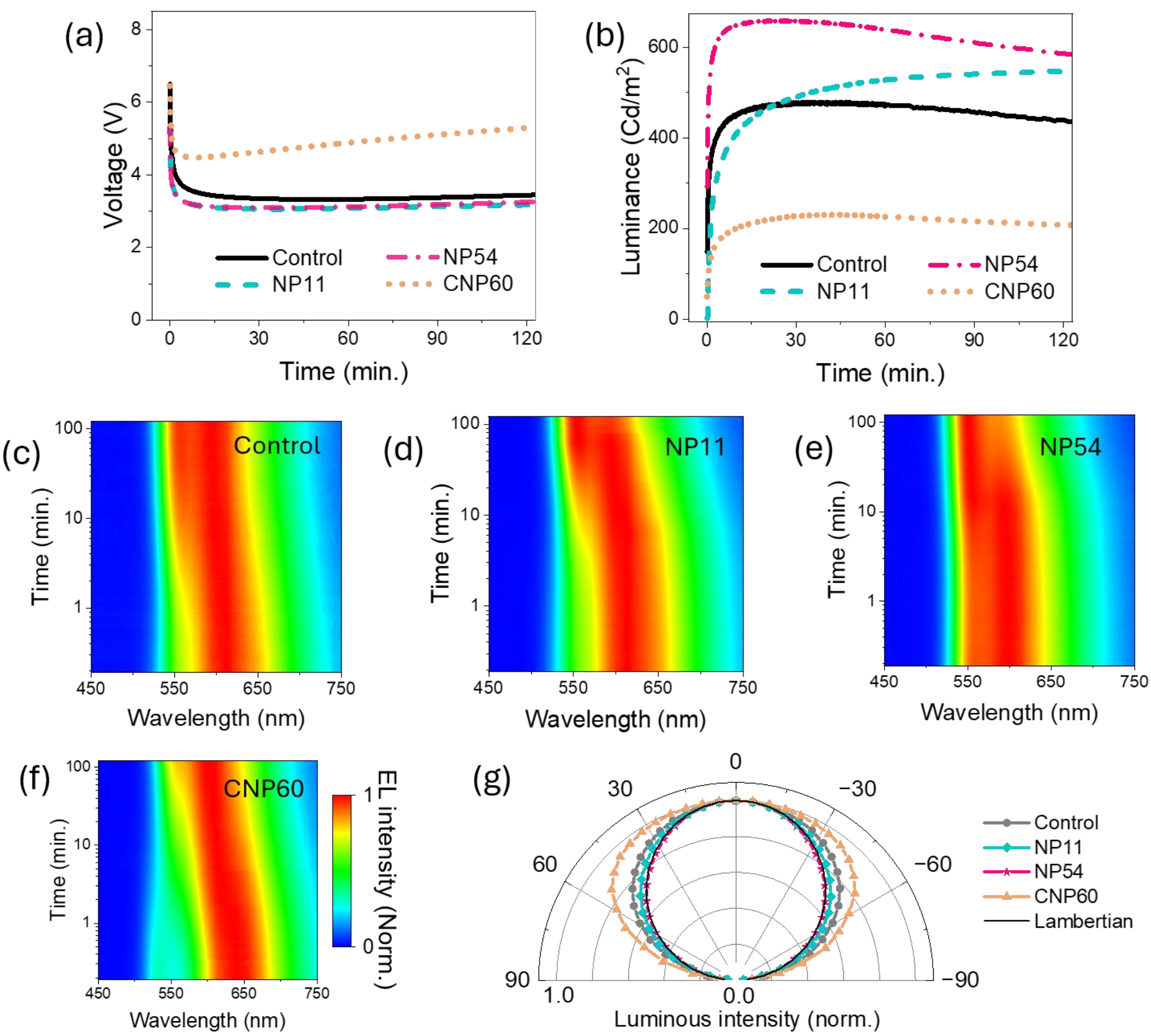

**Figure 2. Measured LEC performance.** The time evolution of (a) the voltage and (b) the luminance for the four LECs when driven by a constant current density of 25 mA $cm^{-2}$. The EL intensity as a function of wavelength and time for (c) the control-LEC, (d) the NP11-LEC, (e) the NP54-LEC, and (f) the CNP60-LEC. (g) The normalized EL intensity as a function of viewing angle at steady state. For reference, the corresponding data for an ideal Lambertian emitter are included.

Figs. 2c-f show the measured transients of the normalized electroluminescence (EL) spectrum for the four LEC devices (each device is identified in the upper right inset), while snapshots of the EL spectrum are presented in Supplementary Fig. 6. The consistent observations are that both the initial (at $t \sim 15$ s) and the final (at $t = 120$ min) EL spectra vary between the four LECs, and that all four LECs exhibit a marked evolution of the EL spectrum with time. Figure 2g presents the normalized EL intensity as a function of viewing angle for the four LECs at $t =$

120 min, while Supplementary Fig. 7 displays the same metric at $t = 10$ min. The viewing-angle dependency of an ideal Lambertian emitter is indicated by the solid black line for reference. The CNP60-LEC (orange curve) exhibits the largest deviation from that of the ideal Lambertian emitter, with a higher relative emission intensity at large viewing angles. In contrast, NP11-LEC and NP54-LEC (light blue and pink curves) exhibit a relatively Lambertian-like dependence, whereas the viewing-angle performance of the control-LEC (grey curve) is in between the CNP60-LEC case and the NP11-LEC and NP54-LEC devices. Since the sole nominal difference between the four LEC devices is the existence and type of Au-NPs at the anode/AM interface, these observations demonstrate that the inclusion of Au-NPs can have significant influence on the overall LEC performance.

Figure 3a displays the simulated luminance as a function of the normalized EZ position in the AM ($\delta_{EZ}$), with $\delta_{EZ} = 0$ corresponding to the anode/AM interface and $\delta_{EZ} = 1$ the AM/cathode interface; details of the simulation procedure are provided in the Methods section. For this specific device architecture, the light emission is strongly reduced when the EZ is located close to either the anode or the cathode, since this is concomitant with destructive interference, severe exciton-polaron quenching[27], and strong non-radiative energy transfer to the electrodes[24]. The peak luminance is instead attained when the EZ is positioned near the center of the AM at $\delta_{EZ} \approx 0.5$. Figure 3b provides additional information in the form of the simulated dependency of the EL spectrum on the EZ position, with the two overlaid EL spectra being determined with EZ positions of $\delta_{EZ} = 0.2$ and $\delta_{EZ} = 0.5$. This strong dependency of the EL intensity and spectrum on the EZ position is useful since it allows for accurate determination of the EZ position in the AM with the aid of the procedure presented in Refs. [28,29].

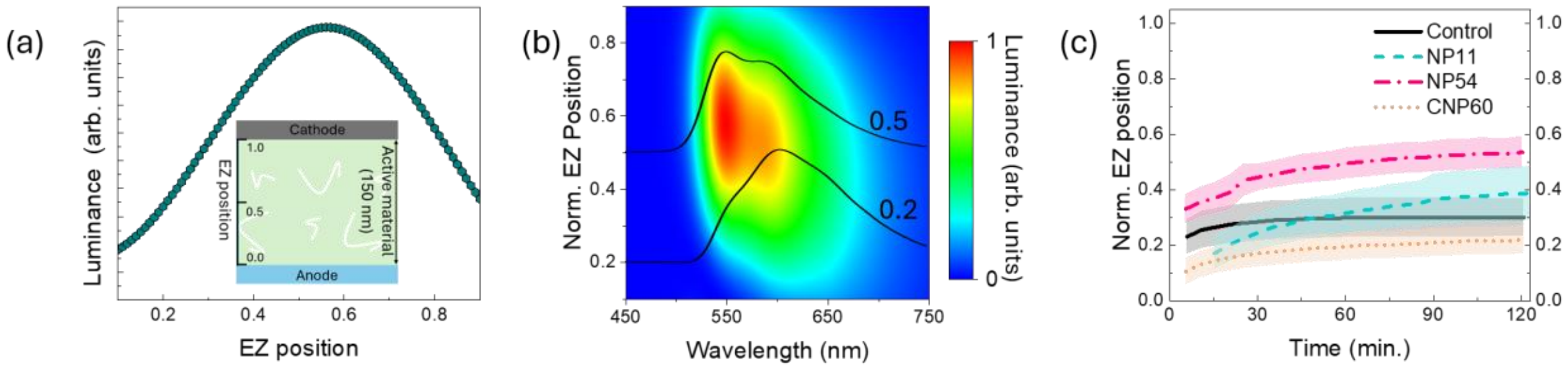

**Figure 3. Simulations and determination of EZ position.** (a) The simulated luminance as a function of the normalized EZ position in the AM, with 0 corresponding to the anode/AM interface, and 1 the AM/cathode interface. (b) The simulated luminance as a function of both the emission wavelength and the EZ position. The two overlaid EL spectra are derived for normalized EZ positions of 0.2 and 0.5. (c) The time evolution of the normalized EZ position in the AM interelectrode gap, as derived from a combination of measurements and simulations.

Figure 3c presents the time evolution of the EZ position in the AM for the four LECs. In the control-LEC, the EZ first forms at $\delta_{EZ} \approx 0.2$ and thereafter migrates slightly to reach $\delta_{EZ} \approx 0.3$ at $t = 120$ min. For the CNP60-LEC, the EZ transient is shifted towards the anode, starting at $\delta_{EZ} \approx 0.1$ and moving to $\delta_{EZ} \approx 0.2$ at steady state. The overall EZ migration is larger for both the NP11-LEC and NP54-LEC, with the EZ shifting from $\approx 0.1$ to $\approx 0.4$ in the former and from $\approx 0.3$ to $\approx 0.55$ in the latter. We mention that the EZ-position results in Fig. 3c are in good qualitative agreement with the simulated luminance in Fig. 3a, and that the measured luminance in Fig. 2b is consistent with the fact that a shift of the EZ towards the center of the AM is concomitant with an increase of the luminance.

We now shift focus to the discussion of our data, and we begin by establishing a few facts. First, the inclusion of metal NPs next to, or into, an emitter matrix can impact the emission properties of the emitters by the Purcell effect, through which the metallic NPs modify the radiative decay rates of nearby emitters[30–35]. However, steady-state and time-resolved measurements show that the photoluminescence intensity and the excited-state lifetime of the SY emitter both remain essentially constant following the NPs incorporation into the AM (see Supplementary Fig. 8). These observations thus rule out plasmonic excitations as a cause for the observed variation in the LEC emission performance.

Second, the inclusion of NPs on a planar electrode surface will induce nanoscale hotspots for the electric field (E-field) next to the NP-modified parts of the electrode surface[36–38], as visualized in the finite-element simulation (using COMSOL Multiphysics[39]) presented in Fig. 4a (details of the simulation procedure are presented in the Methods section). Figs. 4b,c reveal that, although the time for screening the electric field will be shorter around the NPs (due to the initially larger electric field), the concentration of ions required to screen the electrode potential (of 1 V) is the same in both cases.

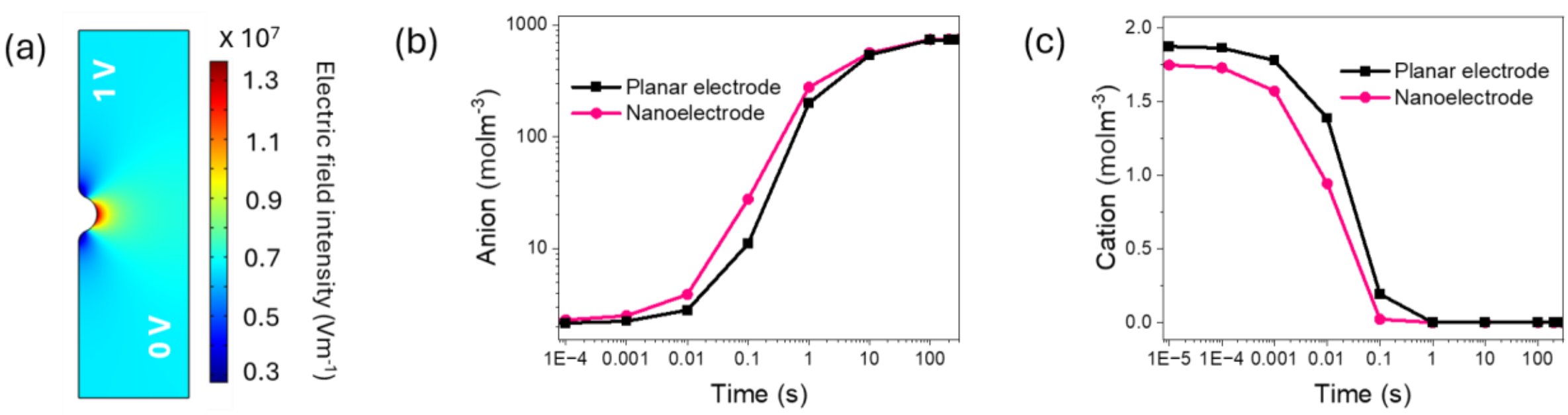

**Figure 4. Simulations of the effects of the NPs on both electric field and ion transport.** (a) Electric field intensity around a nanoelectrode, illustrating localized field profile. Transient anion (b) and cation (c) concentration profiles for a planar anode/AM interface and a nanoelectrode/AM interface, demonstrating faster transport of anions and cations for the nanoelectrode at the early stages of operation.

Given these facts, we start our analysis by focusing on the CNP60-LEC device. The observation that its voltage is increased significantly compared to the other three LECs (Fig. 2a) implies that the electronically insulating ligands on the CNP60 NPs significantly impede the hole injection from the ITO anode. This partial suppression of hole injection shifts the EZ toward the anode (Fig. 3c), which in turn will lead to the observed lowest value for the luminance, as displayed in Fig. 2b. This is consistent with the fact that the coverage of the electrode surface reduces the surface itself.

The origin of the improved emission performance is less straightforward to identify for the NP11-LEC and NP54-LEC devices; or more specifically, why the inclusion of non-capped Au-NPs at the ITO anode causes a shift of the EZ towards the center of the AM (which in turn increases the luminance). Below, we present two possible explanations for the Au-NP-induced EZ shift: (i) the (partial) replacement of ITO with Au as the anode material will decrease the barrier height for hole injection[40] into SY, since the Fermi level of Au is closer to the HOMO of SY[41]. It has been suggested that such a lowering of the barrier height for hole injection will decrease the number of anions in the EDL and by that shift the EZ away from the anode[42,43]; (ii) the steady-state position of the EZ is determined by the ratio between the electron mobility ($\mu_n$) and the hole mobility ($\mu_p$), with a higher value for $\mu_n$ than $\mu_p$ causing the EZ to be located closer to the anode[15,44]. The mobility of conjugated polymers, such as SY, is highly sensitive to the local electric field and charge density[45,46], and it is plausible that the inclusion of the Au-NPs will increase $\mu_p$ of SY next to the anode, and thereby shift the EZ away from the anode towards the center of the AM. We note that the inclusion of the larger-sized NP54 NPs induced the largest anodic EZ shift (Fig. 3c), despite these exhibiting a lower surface coverage of the ITO anode, which would then suggest that mechanism (ii) is dominating.

In summary, we have explored the potential of controlling electrochemical doping in LECs through the incorporation of Au-NPs. We show that the inclusion of Au-NPs at the anode/AM interface can modulate the position of the emissive p-n junction for enhanced emission efficiency. This demonstrates the potential of metal NPs as tunable interfacial elements for controlling electrochemical doping without modifying the bulk composition of the AM. Beyond LECs, this interfacial strategy could have broader implications for emerging technologies that rely on coupled ionic–electronic transport and electrochemical doping for their operation, such as neuromorphic, bioelectronic and energy-storage devices.

## Methods

**Au nanoparticles synthesis and characterization.** Synthesis of Au nanoparticles (NPs): NP11 and NP54 colloids were produced using pulsed laser ablation in liquids (PLAL). A Nd:YAG laser system (Surelite Continuum II, pulse width 5 nm and repetition rate 10 Hz) operating at its fundamental wavelength (1064 nm) was employed, delivering laser fluences in the range of 0.6–1 J/cm². The beam was focused onto a gold target of 99.95% purity (Alfa Aesar) using a lens with a focal length of 10 cm. For samples NP11, the ablation process was carried out in an aqueous sodium chloride solution (10 mM) adjusted to pH 9. Sample NP54 was obtained under identical PLAL conditions using Milli-Q water as the liquid medium (pH7). After synthesis, this colloid was irradiated with laser (wavelength 532 nm) for 20 minutes with a lower fluence of 30 mJ/cm². The NPs' size was tuned by modifying the ablation medium and by introducing an additional laser irradiation step. CNP60 colloids dispersed in water were obtained from BBI Solutions and used as received. The size distribution of NP11 and NP54 is presented in Supplementary Fig. 1, were characterized in collaboration with the Umeå Centre for Electron Microscopy (UCEM). Transmission electron microscopy (TEM) analysis was conducted using the FEI Talos L 120C microscope, operating at 120 kV. To prepare the samples, a few drops of colloidal NPs were placed on a lacey carbon copper grid, making them suitable for TEM imaging. NP size distributions were estimated by analyzing numerous TEM micrographs with ImageJ software. The extinction coefficient of NPs was measured with Perkin UV-Vis spectrometer by dispersing in the solution and presented in Supplementary Fig. 2. It reveals the presence of localized surface plasmon resonance peak in the regime of approximately 520 nm and a red shift in the resonance peak was observed with the increase in size of NPs.

**Device fabrication.** The electroluminescent organic OSC was a phenyl-substituted poly(paraphenylene vinylene) conjugated copolymer termed "Super Yellow" (SY) (Merck, Darmstadt, DE), the salt was $KCF_3SO_3$ (Sigma Aldrich, USA), and the ion transporter was TMPE, which was end-capped with hydroxyls (TMPE-OH, $M_n = 1014$ g mol$^{-1}$, Sigma Aldrich, USA). The salt and the ion transporters were dried in a vacuum oven at 150 °C and 50 °C, respectively, for 12 h before the preparation of inks, while SY was used as received. The active-material (AM) constituents were separately dissolved in in cyclohexanone (Sigma Aldrich, USA) in the following concentrations: $KCF_3SO_3$ and TMPE-OH: 20 g L$^{-1}$; SY: 8 g L$^{-1}$; The active-material ink was prepared by mixing these master inks in a solute mass ratio of SY:TMPE-OH:$KCF_3SO_3$ = 1:0.2:0.06, The transparent indium-tin-oxide (ITO) coated glass substrate (145 nm, $R_s = 20\ \Omega\square^{-1}$, thin film devices) was cleaned by sequential ultrasonic treatment in detergent (Extran MA 01, Merck), deionized water, acetone, and isopropanol. The AM ink was spin-coated on the ITO-coated substrate for 60 s, with a spin speed of 1300 rpm and acceleration of 800 rpm/sec, and thereafter dried at 70 °C for at 1 h. The dry thickness of the AM ($d_{AM}$) was measured with a stylus profilometer (DektakXT, Bruker) and was found to be approximately 150 nm. The reflective Al top electrode (thickness = 150 nm) was deposited by thermal evaporation under vacuum ($p < 8 \times 10^{-6}$ mbar) through a shadow mask. The overlap of the transparent ITO anode and the reflective Al cathode defined four independent 2 × 2 mm$^2$ LEC devices on each substrate. For ambient environment measurements, LECs were protected by attaching a thin glass lid on top of the Al cathodes, using a UV-curable epoxy (Ossila). All the NPs were deposited by spin-coating the NPs on ITO-coated glass substrates for 60 s with a spin speed of 1000 rpm and acceleration of 500 rpm/sec. After, the films were dried on a hot plate at 120° for 15 min. The scanning electron microscopy (SEM) images of NPs on ITO substrates are presented in the Supplementary Fig. 3. Since LEC devices comprise an active layer on top of ITO, we needed to check the presence of NPs after the deposition of the active layer. Supplementary Figs. 4a-c show the SEM images of the scratched active layer. Supplementary Figs. 4b and 4c reveal the presence of NPs on the ITO surface after removal of the active layer. Further, the thickness of the active layer after depositing the NPs was measured using a DEKTAK profilometer, Supplementary Fig. 4d. We found that depositing NPs does not change the active layer thickness, which is found to be approximately 150 nm for both control and Au NP incorporated devices. Supplementary Fig. 4d also shows spikes of approximately 20-30 nm for NP11, both on the ITO and active layer surfaces. This observation is in line with the high surface coverage of NP11, observed in the SEM in Supplementary Fig. 3a, and suggests that the active layer conformally coats the NP11.

**Device characterization.** The LECs were driven by a constant current density of 25 mA cm$^{-2}$, with the compliance voltage set to 21 V, and with ITO biased as the positive anode. A source-measure unit (Keithley 2400) supplied the current and recorded the corresponding voltage. The current–voltage characterization of the non-encapsulated devices was performed in a $N_2$-filled glove box ($O_2$ and water content < 1 ppm).

The angle-resolved EL intensity and spectrum of the LECs were measured with a custom-built, automated spectro-goniometer setup. The device under study was placed in a connection jig, which aligned the emission area of the device with the rotation axis of a stepper motor. This rotation defined the viewing angle of the device, with 0° corresponding to the forward emission. The viewing angle was varied between −86.4° and 86.4° in steps of 5.4° and controlled with a Python-based virtual instrument. A fraction of the device emission was collected by a collimating lens (Ø = 7.2 mm, F230 SMA-A, Thorlabs, Germany) positioned 75 mm away from the device, resulting in a small and constant solid collection angle (Ω) of 0.007 sr. An optical fiber delivered the collected light to a CCD-array spectrometer (Flame-S, OceanOptics).

**Setfos simulations.** Modelling of outcoupled light from LEC is performed using the commercial package Setfos (Version 6.0, Fluxim AG, Switzerland). The simulated device architecture is similar to the experimentally fabricated devices and is structured as follows: Air (semi-infinite) / reflective top cathode (100 nm) / SY (150 nm) / ITO (145 nm) / Glass (0.75 mm) / Air (semi-infinite).
In these simulations, the exciton generation profile is approximated by a delta distribution in order to reduce computational complexity. The dipole anisotropy parameter is fixed at a=0.05[47]. The optical constants of SY are used to represent the refractive index of the active material[48] and optical constants of the electrode materials are obtained from the Setfos database. The emission spectrum of SY is taken from the experimentally measured photoluminescence spectrum of a 15 nm thin film of SY. The position of the emission zone (EZ) maximum is scanned from 0.1 to 0.9 along the normalized electrode distance, where 0 denotes the anode and 1 the cathode, with increment of 0.01.

**FEM simulations.** Ion transport was modeled using the Transport of Diluted Species module based on the Nernst–Planck equation:

$$\frac{\partial c_i}{\partial t} + \nabla \cdot J_j + u \cdot \nabla \cdot c_i = 0 \qquad (1)$$

$$J_i = -D_i \nabla c_i - z_i u_{m,j} F c_i \nabla V \qquad (2)$$

$$u_{m,j} = \frac{D_i}{RT} \qquad (3)$$

where $c_i$ is ionic concentration (initial uniform concentrations were set to 2 mol $m^{-3}$ for both cation and anion), $D_i$ is the diffusion coefficient ($D_{cation} = 1 \times 10^{-16}$ $m^2$ $s^{-1}$, $D_{anion} = 1 \times 10^{-16}$ $m^2$ $s^{-1}$), $z_i$ is charge number (+1 and -1 for cation and anion, respectively), F is Faraday's constant, R is the gas constant, T = 298 K, convection velocity u taken as zero[15,49]. The electric field used to drive ion migration was obtained from the electrostatic model with V= 1 V.

**Acknowledgements**
The authors acknowledge Umeå Centre for Electron Microscopy (UCEM) and the National Microscopy Infrastructure (NMI), for instrument access and technical support. N. M. and A. K. P. acknowledge financial support from the Swedish Research Council (Grants No. 2021-05784 and No. 2025-04734), the Knut and Alice Wallenberg Foundation (Grant No. 2023.0089), and Kempestiftelserna (WISE-UmU/Kempe-program, Grant No. JCSMK 23-198). A.K. acknowledges funding from the European Union (HORIZON MSCA 2023 PF, acronym UNID, grant number 101150699). L.E. acknowledges financial support from the European Union through an ERC Advanced Grant (ERC, InnovaLEC, 101096650). L.E., J.R R. and A.K. acknowledge financial support from the Swedish Research Council (2019-02345 and 2021–04778) and the Wallenberg Initiative Materials Science for Sustainability (WISE) funded by the Knut and Alice Wallenberg Foundation (KAW 2024.0497). G. P. A. acknowledges funding from the European Union Program HORIZON-Pathfinder-Open: 3D-BRICKS (Grant No. 101099125) and from the Swiss State Secretariat for Education, Research and Innovation (SERI) (contract number 23.00075).

**References**

1. Berggren, M. & Malliaras, G. G. How conducting polymer electrodes operate. *Science* **364**, 233–234 (2019).
2. *Iontronics: Ionic Carriers in Organic Electronic Materials and Devices*. (CRC Press, Boca Raton, 2017).
3. Paulsen, B. D., Tybrandt, K., Stavrinidou, E. & Rivnay, J. Organic mixed ionic–electronic conductors. *Nat. Mater.* **19**, 13–26 (2020).
4. Sood, A. *et al.* Electrochemical ion insertion from the atomic to the device scale. *Nat. Rev. Mater.* **6**, 847–867 (2021).
5. Wang, Y. *et al.* Designing organic mixed conductors for electrochemical transistor applications. *Nat. Rev. Mater.* **9**, 249–265 (2024).
6. Tang, S. & Edman, L. Light-Emitting Electrochemical Cells: A Review on Recent Progress. in *Photoluminescent Materials and Electroluminescent Devices* (eds Armaroli, N. & Bolink, H. J.) 375–395 (Springer International Publishing, Cham, 2017). doi:10.1007/978-3-319-59304-3_12.
7. Fan, X. *et al.* Opportunities of Flexible and Portable Electrochemical Devices for Energy Storage: Expanding the Spotlight onto Semi-solid/Solid Electrolytes. *Chem. Rev.* **122**, 17155–17239 (2022).
8. Zheng, L., Zhu, X. & Xiao, K. Photo-iontronics: mechanisms and manipulation for neuromorphic vision. *Iontronics* **2**, (2026).
9. Chen, K. *et al.* Organic optoelectronic synapse based on photon-modulated electrochemical doping. *Nat. Photonics* **17**, 629–637 (2023).
10. Wang, B. *et al.* Organic electrochemical transistors for integrated sensing–memory–computing hardware towards intelligent Internet of Bodies. *Nat. Rev. Electr. Eng.* https://doi.org/10.1038/s44287-025-00234-x (2025) doi:10.1038/s44287-025-00234-x.
11. Scaccabarozzi, A. D. *et al.* Doping Approaches for Organic Semiconductors. *Chem. Rev.* **122**, 4420–4492 (2022).

12. Tang, S. *et al.* Design rules for light-emitting electrochemical cells delivering bright luminance at 27.5 percent external quantum efficiency. *Nat. Commun.* **8**, 1190 (2017).
13. Liu, T. *et al.* Gold-activated persulfate p-doping of organic semiconductors. *Nat. Mater.* https://doi.org/10.1038/s41563-026-02547-0 (2026) doi:10.1038/s41563-026-02547-0.
14. Xiang, L. *et al.* Nanoscale doping of polymeric semiconductors with confined electrochemical ion implantation. *Nat. Nanotechnol.* **19**, 1122–1129 (2024).
15. Ràfols-Ribé, J. *et al.* Controlling the Emission Zone by Additives for Improved Light-Emitting Electrochemical Cells. *Adv. Mater.* **34**, 2107849 (2022).
16. Keene, S. T. *et al.* Hole-limited electrochemical doping in conjugated polymers. *Nat. Mater.* **22**, 1121–1127 (2023).
17. Zhang, Z. *et al.* A colour-tunable, weavable fibre-shaped polymer light-emitting electrochemical cell. *Nat. Photonics* **9**, 233–238 (2015).
18. Tang, S., Pan, J., Buchholz, H. A. & Edman, L. White Light from a Single-Emitter Light-Emitting Electrochemical Cell. *J. Am. Chem. Soc.* **135**, 3647–3652 (2013).
19. Yasuji, K., Sakanoue, T., Yonekawa, F. & Kanemoto, K. Visualizing electroluminescence process in light-emitting electrochemical cells. *Nat. Commun.* **14**, 992 (2023).
20. Slinker, J. D. *et al.* Direct measurement of the electric-field distribution in a light-emitting electrochemical cell. *Nat. Mater.* **6**, 894–899 (2007).
21. Liu, J. *et al.* Fully stretchable active-matrix organic light-emitting electrochemical cell array. *Nat. Commun.* **11**, 3362 (2020).
22. Tang, S., Tsuchiya, Y., Wang, J., Adachi, C. & Edman, L. White light-emitting electrochemical cells based on metal-free TADF emitters. *Nat. Commun.* **16**, 653 (2025).
23. Yen, M.-C. *et al.* All-inorganic perovskite quantum dot light-emitting memories. *Nat. Commun.* **12**, 4460 (2021).
24. Zhang, X. *et al.* Efficiency Roll-Off in Light-Emitting Electrochemical Cells. *Adv. Mater.* **36**, 2310156 (2024).
25. Tang, S. & Edman, L. Quest for an Appropriate Electrolyte for High-Performance Light-Emitting Electrochemical Cells. *J. Phys. Chem. Lett.* **1**, 2727–2732 (2010).
26. Fang, J., Yang, Y. & Edman, L. Understanding the operation of light-emitting electrochemical cells. *Appl. Phys. Lett.* **93**, 063503 (2008).
27. Zhang, X. *et al.* Efficiency Roll-Off in Light-Emitting Electrochemical Cells. *Adv. Mater.* **36**, 2310156 (2024).
28. Lindh, E. M., Lundberg, P., Lanz, T., Mindemark, J. & Edman, L. The Weak Microcavity as an Enabler for Bright and Fault-tolerant Light-emitting Electrochemical Cells. *Sci. Rep.* **8**, 6970 (2018).
29. Lindh, E. M., Lundberg, P., Lanz, T. & Edman, L. Optical analysis of light-emitting electrochemical cells. *Sci. Rep.* **9**, 10433 (2019).
30. Fusella, M. A. *et al.* Plasmonic enhancement of stability and brightness in organic light-emitting devices. *Nature* **585**, 379–382 (2020).
31. Shih, H.-C., Chiou, B.-R. & Su, H.-C. Plasmonic enhancement in electroluminescence from light-emitting electrochemical cells incorporating gold nanourchins. *J. Mater. Chem. C* **4**, 5610–5616 (2016).
32. Gu, X., Qiu, T., Zhang, W. & Chu, P. K. Light-emitting diodes enhanced by localized surface plasmon resonance. *Nanoscale Res. Lett.* **6**, 199 (2011).

33. Zambrana-Puyalto, X. *et al.* Site-selective functionalization of plasmonic nanopores for enhanced fluorescence emission rate and Förster resonance energy transfer. *Nanoscale Adv.* **1**, 2454–2461 (2019).
34. Zambrana-Puyalto, X. *et al.* A hybrid metal-dielectric zero mode waveguide for enhanced single molecule detection. *Chem. Commun.* **55**, (2019).
35. Zambrana-Puyalto, X., Ponzellini, P., MacCaferri, N. & Garoli, D. Förster-Resonance Energy Transfer between Diffusing Molecules and a Functionalized Plasmonic Nanopore. *Phys. Rev. Appl.* **14**, 1 (2020).
36. Liu, M. *et al.* Enhanced electrocatalytic CO2 reduction via field-induced reagent concentration. *Nature* **537**, 382–386 (2016).
37. Cui, C., Jin, R., Jiang, D., Zhang, J. & Zhu, J. Visualization of an Accelerated Electrochemical Reaction under an Enhanced Electric Field. *Research* **2021**, 2021/1742919 (2021).
38. Huang, J.-A. *et al.* SERS discrimination of single DNA bases in single oligonucleotides by electro-plasmonic trapping. *Nat. Commun.* **10**, 5321 (2019).
39. COMSOL AB, "COMSOL Multiphysics® v. 6.3," https://www. comsol.com (2025), stockholm, Sweden.
40. Keller, K. R. *et al.* Ultrafast Thermionic Electron Injection Effects on Exciton Formation Dynamics at a van der Waals Semiconductor/Metal Interface. *ACS Photonics* **9**, 2683–2690 (2022).
41. Park, J. H. *et al.* Polymer/Gold Nanoparticle Nanocomposite Light-Emitting Diodes: Enhancement of Electroluminescence Stability and Quantum Efficiency of Blue-Light-Emitting Polymers. *Chem. Mater.* **16**, 688–692 (2004).
42. Kirch, A. *et al.* Impact of the Electrode Material on the Performance of Light-Emitting Electrochemical Cells. *ACS Appl. Mater. Interfaces* **17**, 5184–5192 (2025).
43. Xu, J. *et al.* Challenging Conventional Wisdom: Finding High-Performance Electrodes for Light-Emitting Electrochemical Cells. *ACS Appl. Mater. Interfaces* **10**, 33380–33389 (2018).
44. Diethelm, M. *et al.* The Dynamic Emission Zone in Sandwich Polymer Light-Emitting Electrochemical Cells. *Adv. Funct. Mater.* **30**, 1906803 (2020).
45. Blom, P. W. M., De Jong, M. J. M. & Van Munster, M. G. Electric-field and temperature dependence of the hole mobility in poly(p-phenylene vinylene). *Phys. Rev. B* **55**, R656–R659 (1997).
46. Pasveer, W. F. *et al.* Unified Description of Charge-Carrier Mobilities in Disordered Semiconducting Polymers. *Phys. Rev. Lett.* **94**, 206601 (2005).
47. Ràfols-Ribé, J., Hänisch, C., Larsen, C., Reineke, S. & Edman, L. In Situ Determination of the Orientation of the Emissive Dipoles in Light-Emitting Electrochemical Cells. *Adv. Mater. Technol.* **8**, 2202120 (2023).
48. Lanz, T., Lindh, E. M. & Edman, L. On the asymmetric evolution of the optical properties of a conjugated polymer during electrochemical p- and n-type doping. *J. Mater. Chem. C* **5**, 4706–4715 (2017).
49. Ràfols-Ribé, J. *et al.* Pinpointing the Dynamic $p$ - $i$ - $n$ Junction. *PRX Energy* **4**, 033015 (2025).

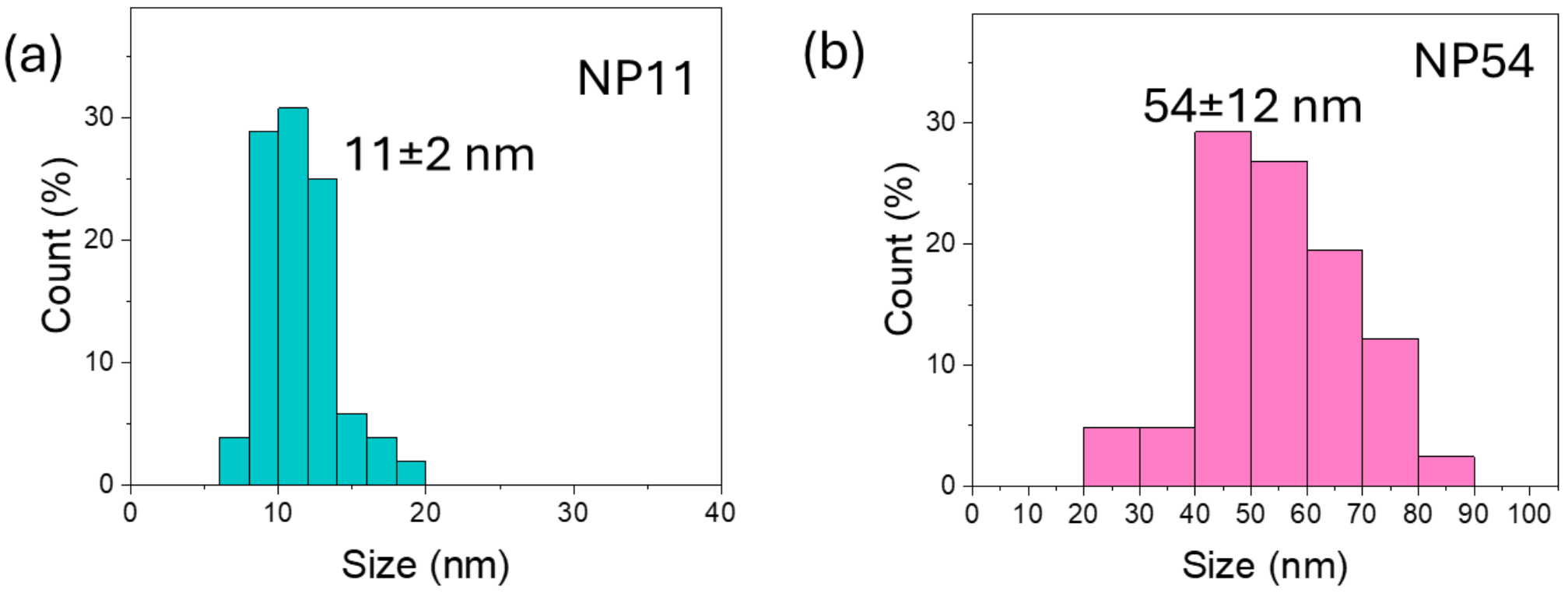

**Supplementary Figure 1.** Size distribution of Au NPs, obtained by TEM measurements for (a) NP11, (b) NP54.

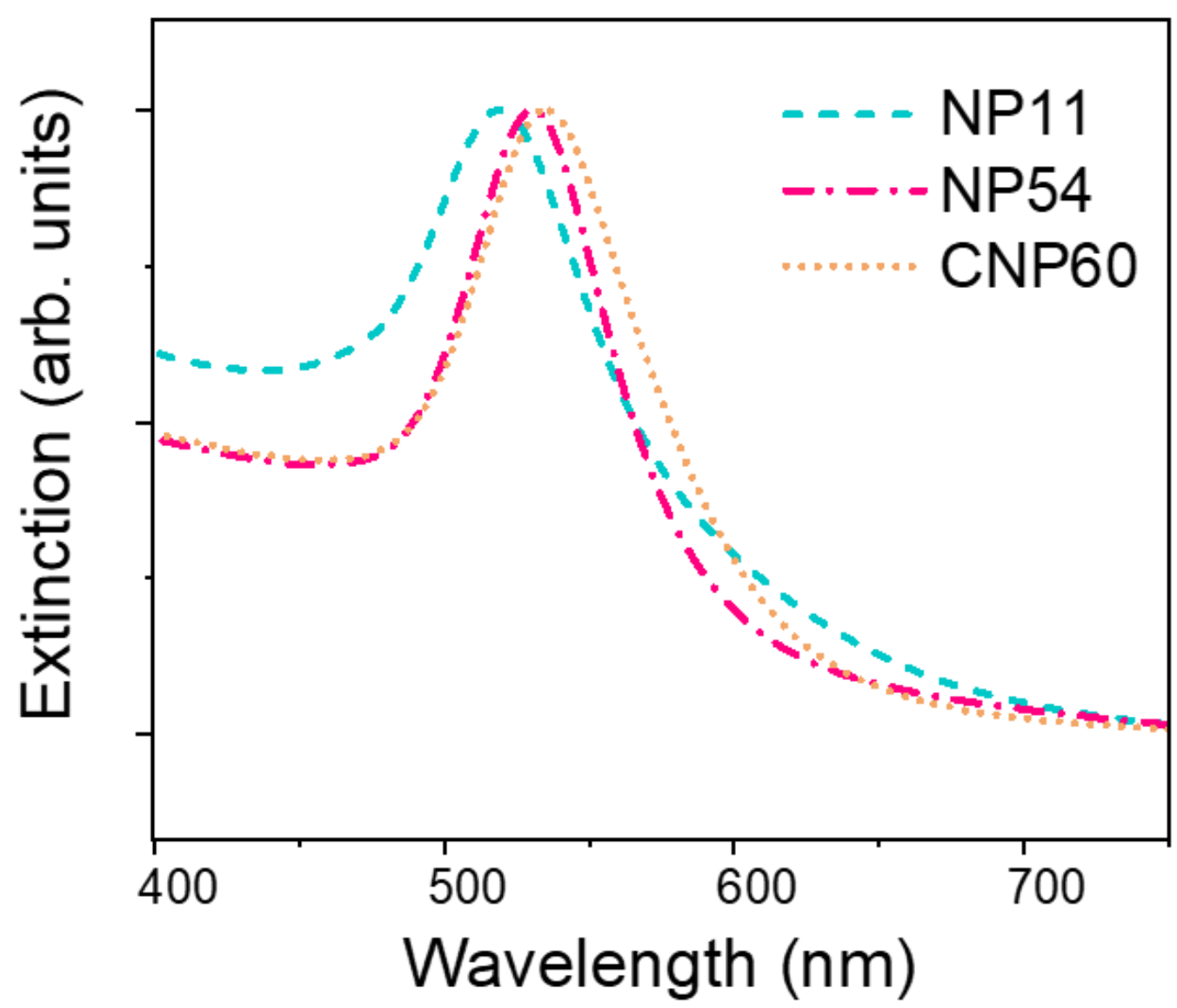

**Supplementary Figure 2.** Extinction spectra of Au NPs in the solution.

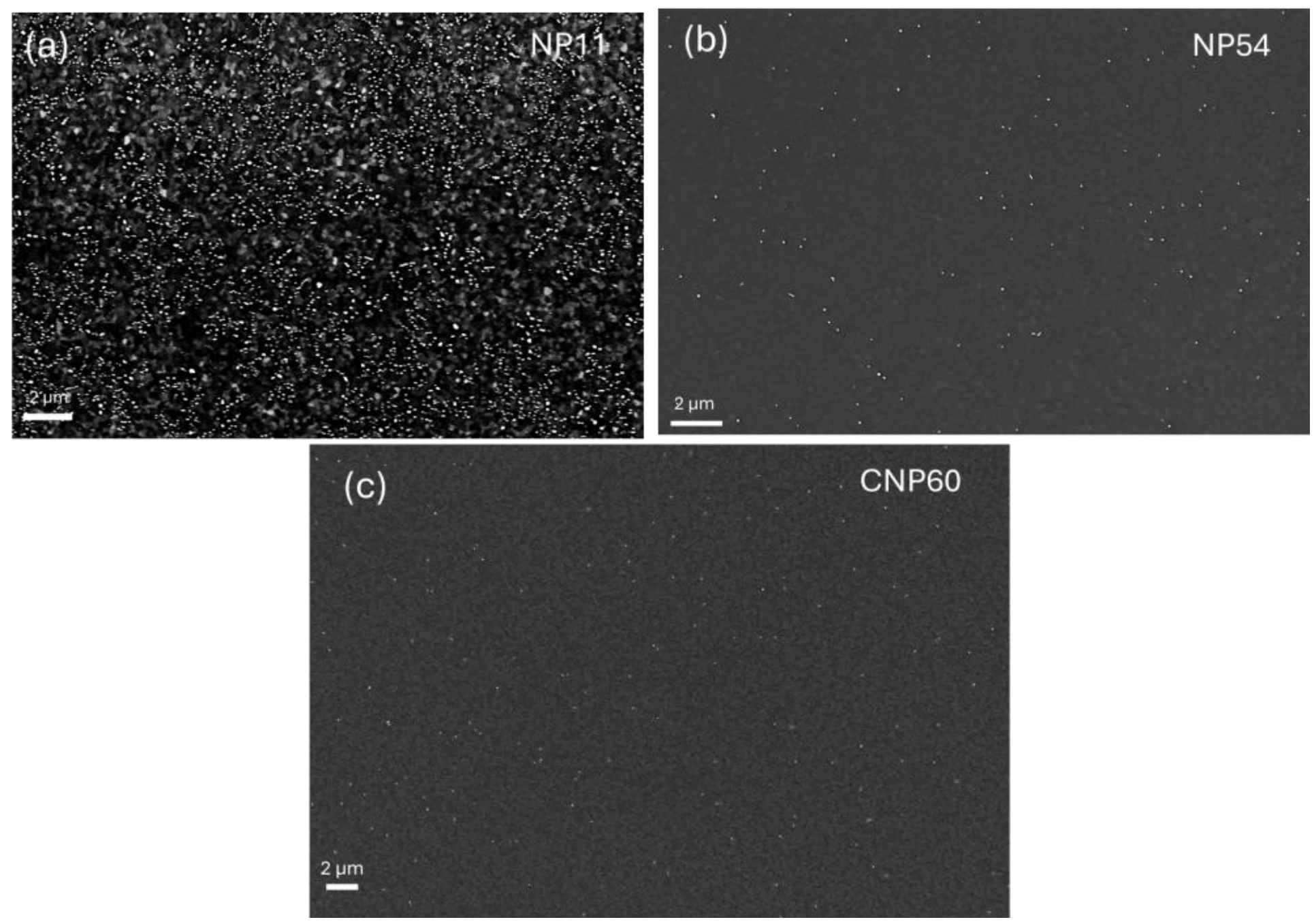

**Supplementary Figure 3.** SEM images of spin coated NPs (white dots) on ITO coated glass substrate: (a) NP11, (b) NP54, and (c) CNP60.

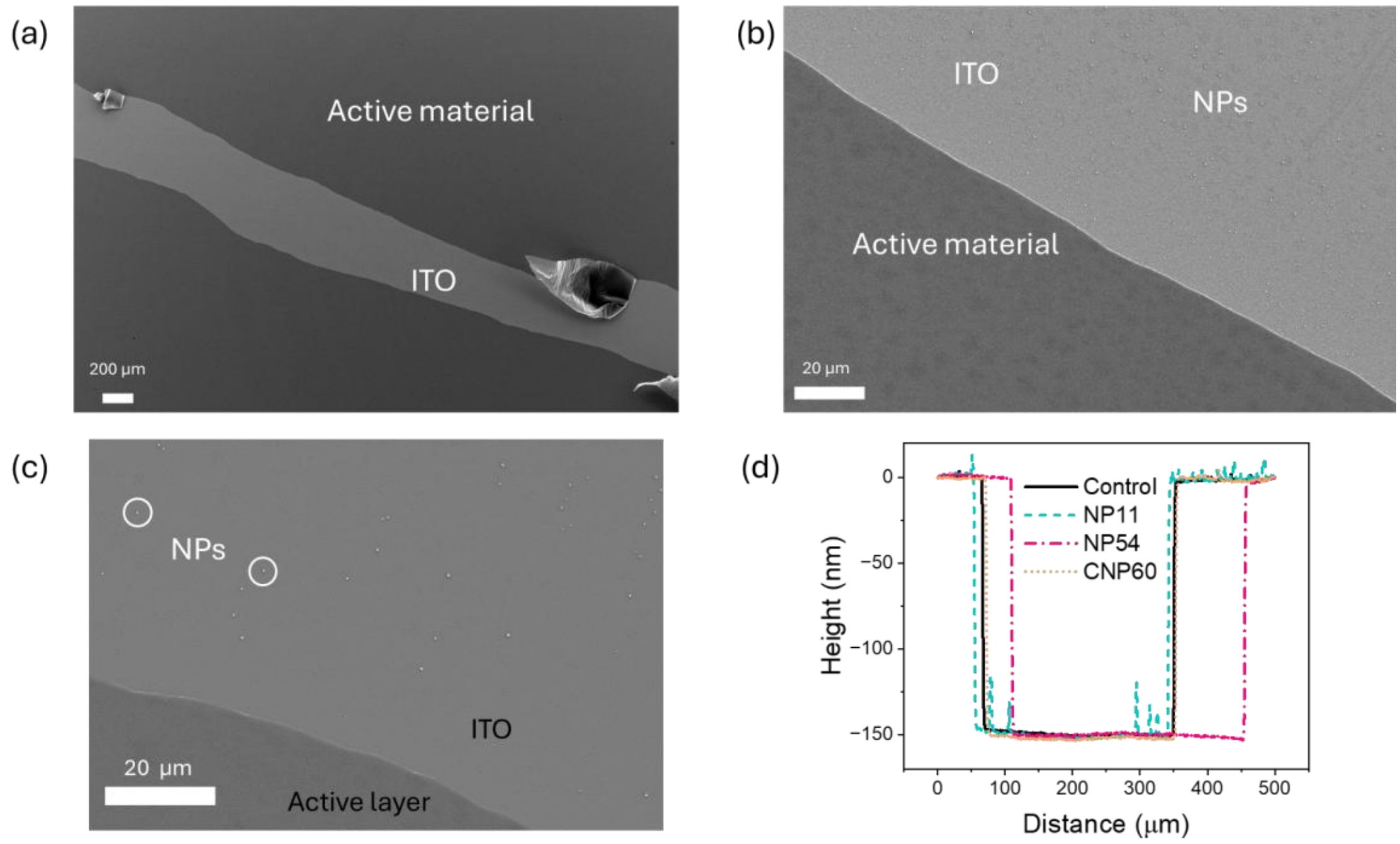

**Supplementary Figure 4.** (a) SEM image of active layer coated on NP11 deposited ITO substrate, active layered peeled off from the center of the image. (b, c) Zoomed SEM images showing NPs (while dots) on top of ITO. (d) The active-layer thickness of control and NP-LECs.

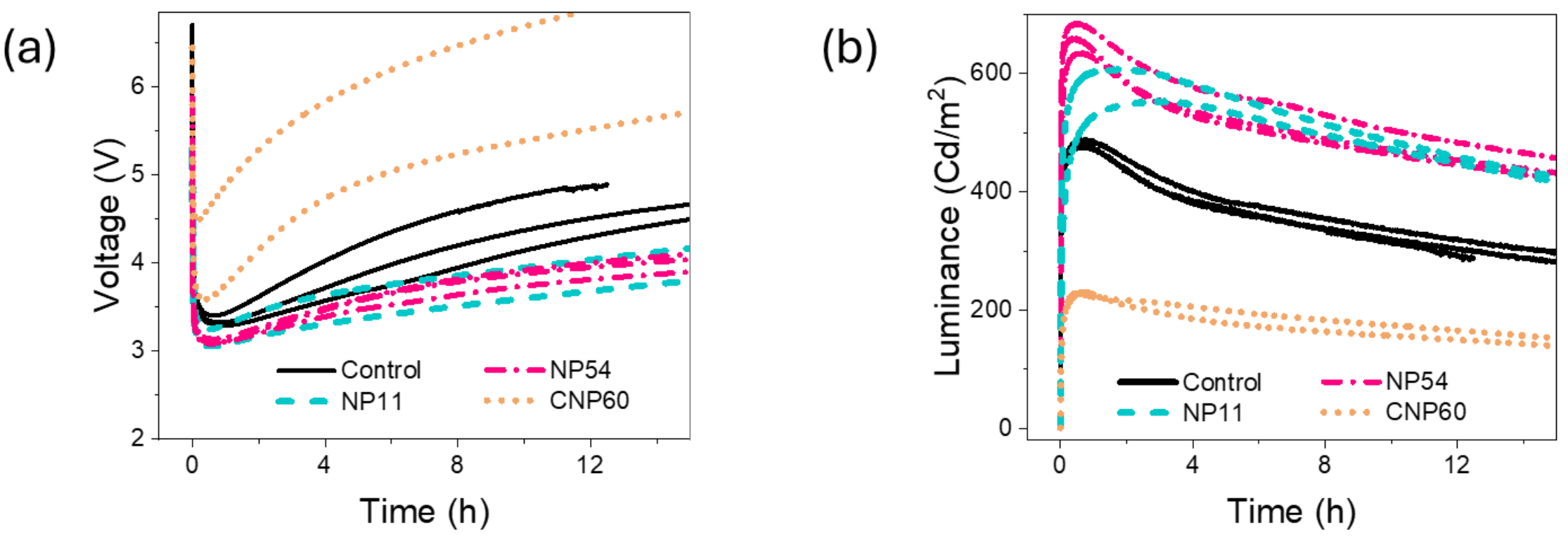

**Supplementary Figure 5**. Time evolution of (a) voltage (b) luminance for control- and NP-LECs driven by under a current density of 25 mA cm$^{-2}$.

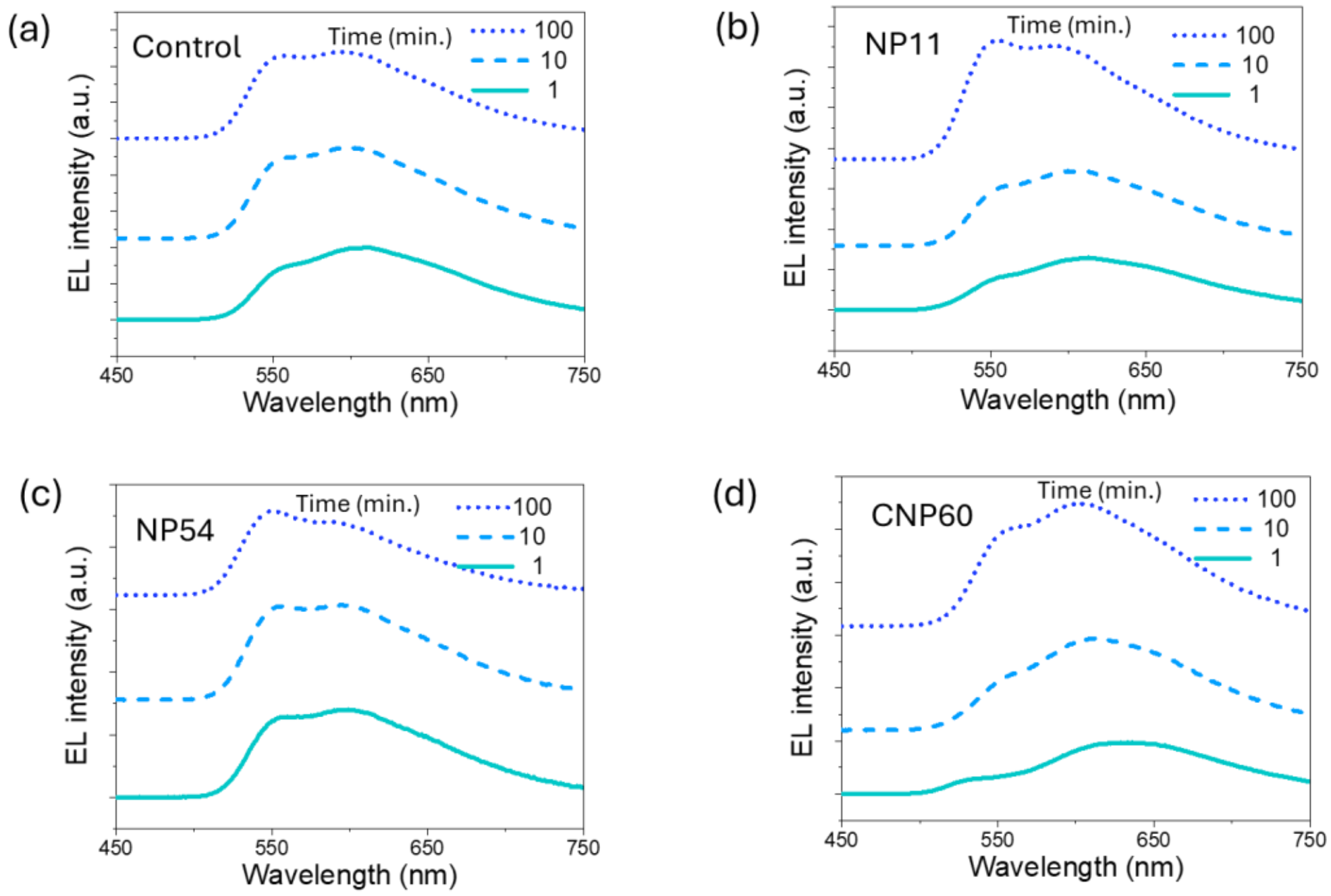

**Supplementary Figure 6.** EL cross section at selected times (a) control-LEC (b) NP11-LEC (c) NP54-LEC (d) CNP60-LEC, showing distinct spectral profiles for control- and NP-LECs corresponding to different EZ positions.

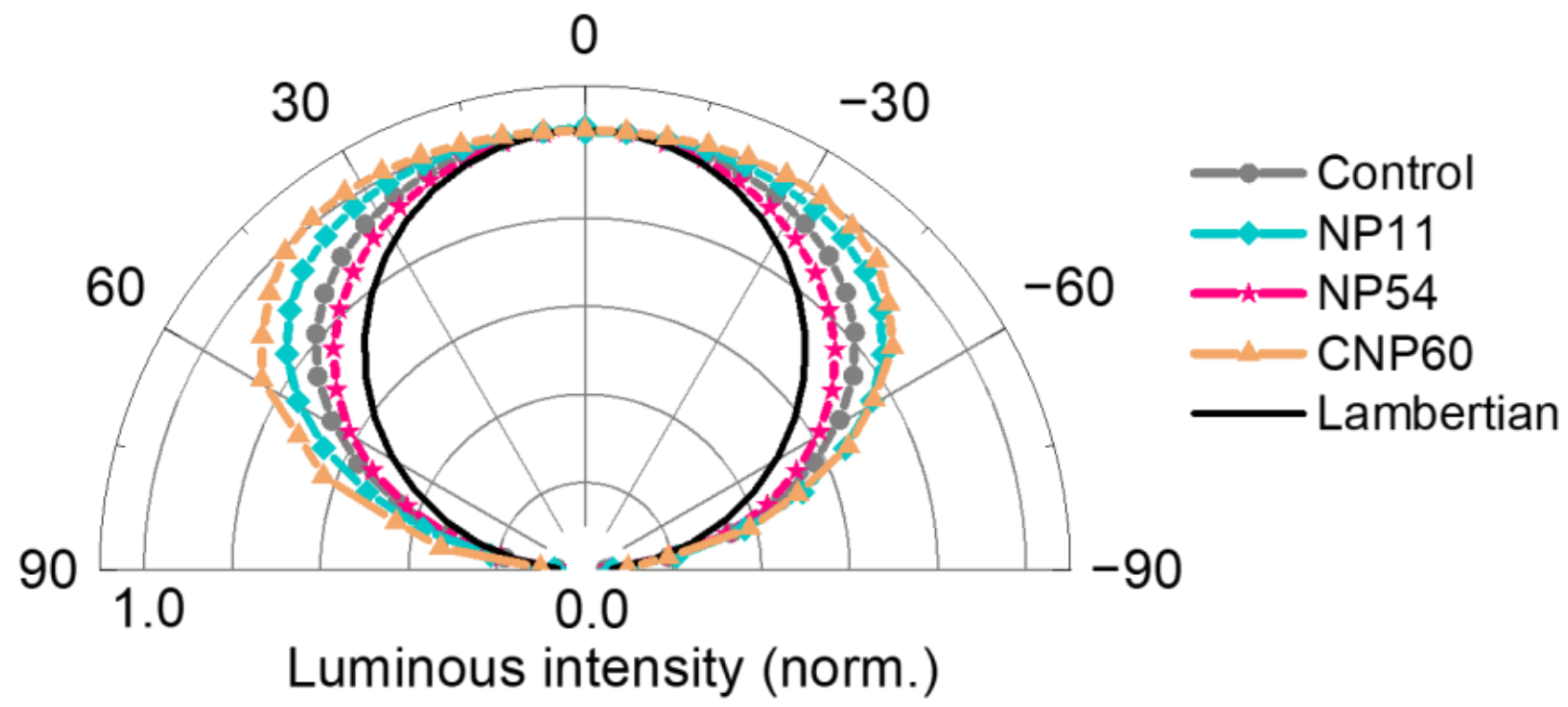

**Supplementary Figure 7.** Polar plots of normalized EL intensity versus emission angle for control- and NP-LECs in the early stage of operation (at t= 10 min).

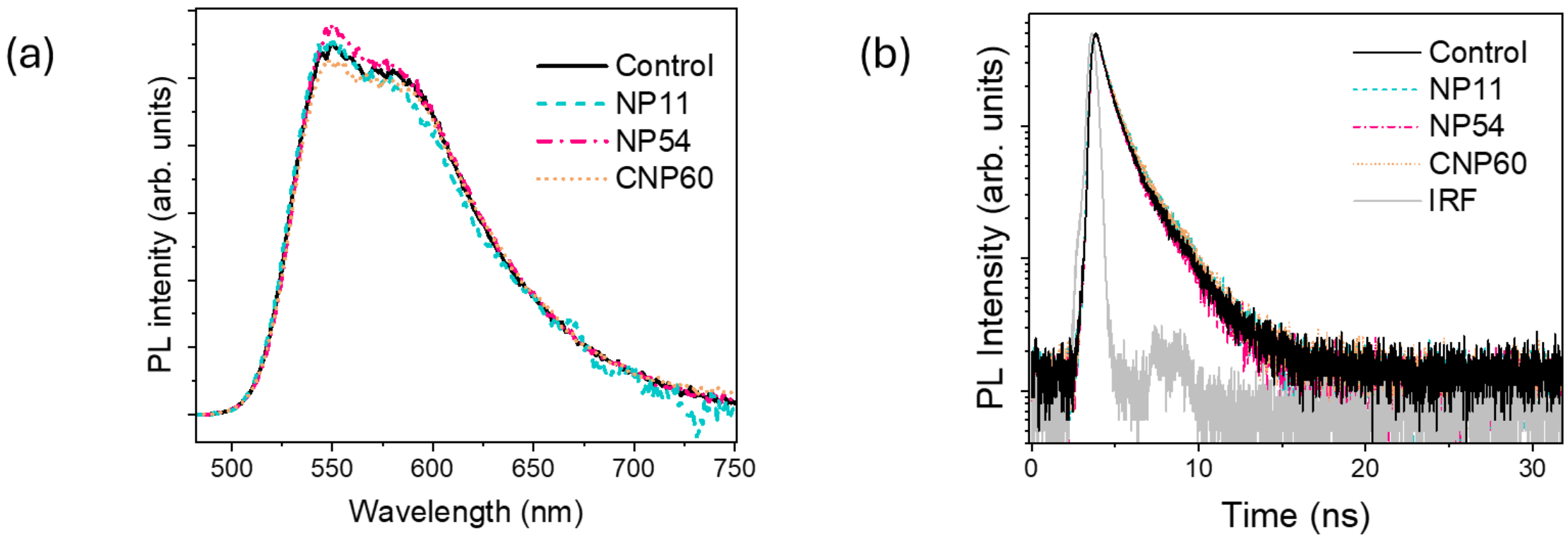

**Supplementary Figure 8**. (a) Steady state photoluminescence of control and NP-LECs (b) Lifetime of photoluminescence (measured with time-correlated single-photon counting (TCSPC) spectroscopy) for of control and NP-LECs.